\documentclass[10pt,twocolumn,letterpaper]{article}
\usepackage[pagenumbers]{nbtf}

\usepackage{cuted} %

\usepackage{currfile} %

\usepackage{caption} %

\newif\ifshowchanges \showchangesfalse
\newcommand{\changed}[1]{\ifshowchanges{\color{blue}#1}\else#1\fi}
\usepackage{tikz}
\usepackage{placeins}
\usetikzlibrary{positioning,arrows.meta,calc}
\definecolor{linkblue}{rgb}{0.21,0.49,0.74}
\usepackage[pagebackref,breaklinks,colorlinks,allcolors=linkblue]{hyperref}

\title{One Code per Material: Neural Representation and Generation of Measured BTFs}
\author{Fazilet Gokbudak \quad Canberk Baykal \quad Chenliang Zhou \quad Rafal Mantiuk \quad Cengiz Oztireli\\ University of Cambridge}

\begin{document}
\maketitle
\begin{strip}
\centering
\includegraphics[width=\linewidth]{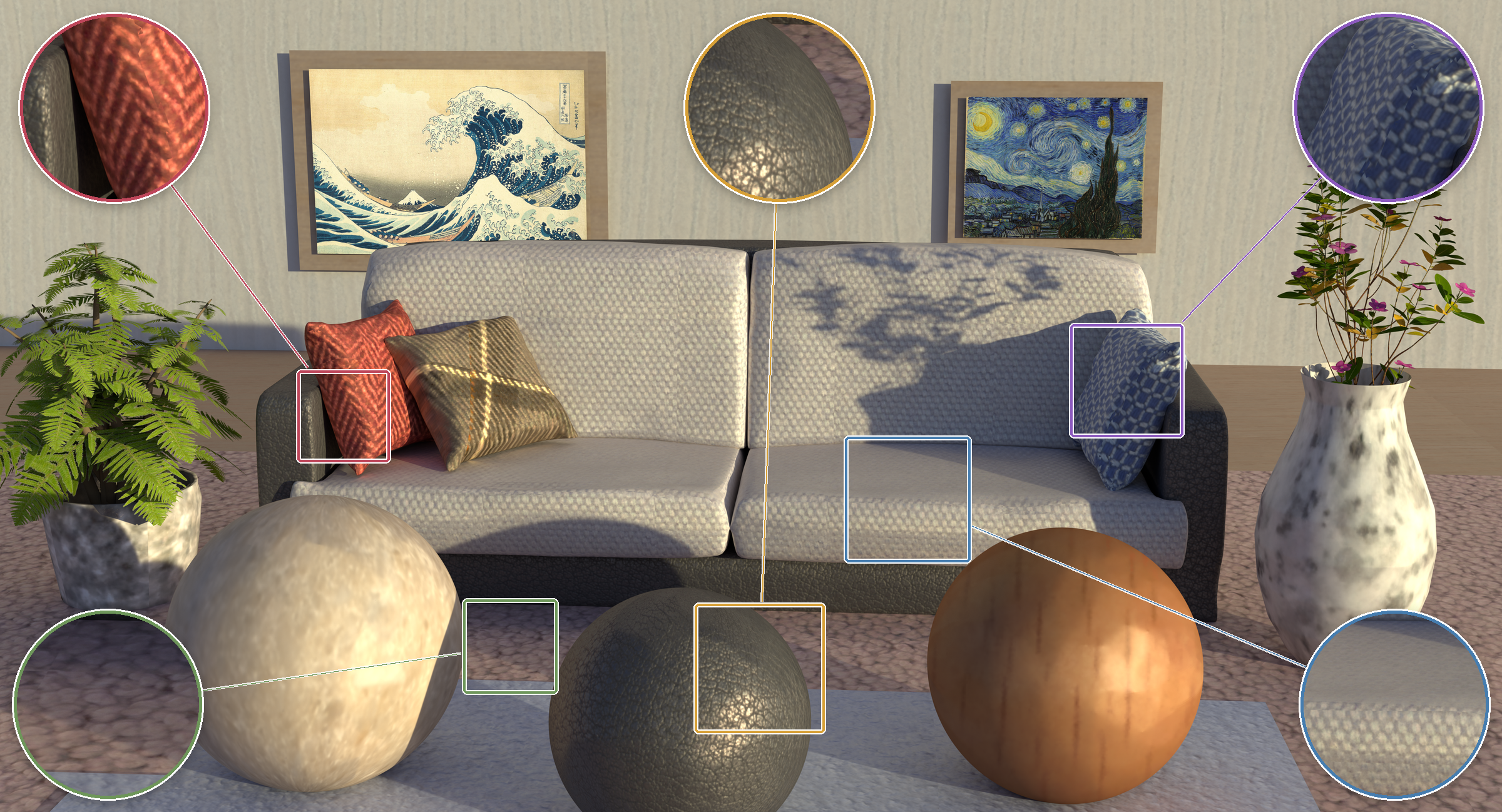}
\captionsetup{hypcap=false}
\captionof{figure}{\textbf{\changed{From measured to sampled BTFs in one scene,}} rendered with Mitsuba~3~\cite{jakob2022mitsuba3}.
\changed{Every textured surface is a BTF reproduced or generated by our models: the per-material networks and the compact
model represent measured materials, and the flow model samples further ones; only the paintings and plants use standard
textures. Close-ups: towards grazing angles, the relief of leather, fabric and carpet keeps
the measured shading and shadowing between neighbouring texels, an effect a per-texel SVBRDF cannot
represent~\cite{dana1999reflectance}.}
}
\label{fig:teaser}
\captionsetup{hypcap=true}
\vspace{-12pt}
\end{strip}

\begin{abstract}
Measured bidirectional texture functions (BTFs) capture the full appearance of real materials---including
parallax, self-shadowing and inter-reflections that analytic SVBRDFs cannot express---but each measurement is
billions of values, and every new material requires a new capture. Existing neural BTF representations
encode each material as large per-texel latent maps rather than as a point in a shared
space from which new materials could be sampled.
We present neural representations of measured BTFs at two operating points. A per-material coordinate network
outperforms the recent neural BTF of Xia \etal at matched size (+3.2\,dB PSNR) and all prior baselines on every metric. A compact model represents
each material by a single 256-dimensional code that conditions a shared decoder; it exceeds the PSNR (+1.2\,dB) and colour accuracy of the per-texel latent maps of Rainer \etal with 256 numbers
per material instead of millions, and one decoder shared by all materials.
Because each material is a single point in this code space, measured BTFs can be sampled: a class-conditional
flow over the codes, trained on the 84 materials enlarged with colourways and hue rotations, generates materials
that render and relight as full BTFs.
\end{abstract}

\section{Introduction}
\label{sec:intro}

Real materials owe much of their appearance to structure below the scale of a surface: the loops of a carpet,
the crack network of leather, the weave of a fabric. Under changing light and view these structures cast shadows,
occlude each other and shift with parallax. Bidirectional texture functions (BTFs)~\cite{dana1999reflectance}
capture this appearance directly by measuring a planar sample under many light and view directions, and
measured BTF datasets such as UBO2014~\cite{weinmann2014material} remain the reference for realistic
material appearance. Analytic SVBRDFs, such as the metallic--roughness model of most material
generators~\cite{burley2012physically,karis2013real}, cannot represent such complex effects, which give real materials their realism (\cref{sec:exp_recon}).

The price of this fidelity is size. A single UBO2014 material is 400$\times$400 texels under
151$\times$151 light--view pairs, about $10^{10}$ values, and creating a new material requires a new capture.
Neural BTF representations reduce the storage by decoding per-texel latent features with a small network:
latent maps~\cite{rainer2019neural,rainer2020unified}, feature textures with neural
offsets~\cite{kuznetsov2021neumip,xia2026neural}, or feature planes~\cite{fan2023neural}.
These representations reconstruct well, but store a feature vector \emph{per texel} and are not designed to sample
\emph{new} materials. Generative material models, on the
other hand, learn from large synthetic corpora or from single flash photographs and produce parametric
SVBRDFs~\cite{henzler2021generative,zhou2023photomat,munkberg2026texture} or neural materials fitted to synthetic
data~\cite{raghavan2025generative}; none of them learns from measured BTFs, of which only about a hundred
are publicly available.

We bring the two together: a representation of measured BTFs that is \changed{accurate and compact, and whose codes can be sampled}. We achieve
this in three steps (\cref{fig:overview}). A per-material coordinate network
gives the most accurate representation. A compact model replaces the
per-material network by a single global code per material and a shared decoder, so a material is described by 256 floats.
Because each material is a point in this low-dimensional space, \changed{the codes also lend themselves to generation,
which we explore with a flow model~\cite{liu2022flow,lipman2023flow} trained on the codes: it samples materials, and}
the same decoder renders them as full BTFs with measured angular behaviour.

Measured BTF datasets are inherently small, with UBO2014 containing only twelve materials per class, making it
difficult to learn a distribution of material appearances. To address this limitation, we augment the
training set through colourway transfer, combining the measured structure of one material with the colour
characteristics of another material from the same class. This
expands the dataset from 84 to 886 materials, so that the generator learns colour variation over the measured
structures.

In summary, we contribute:
\begin{itemize}
  \item a per-material neural BTF that outperforms all neural BTF baselines at comparable size (2.6--3.4M values per
        material);
  \item a compact model with one 256-float code per material and a shared decoder that exceeds the PSNR and colour
        accuracy of the per-texel latent maps of Rainer \etal~\cite{rainer2020unified} with 14$\times$ fewer values per
        material, even counting the shared decoder;
  \item a code space in which measured BTFs can be sampled: a class-conditional flow over the codes \changed{learns the distribution of} 84
        captures enlarged with colourways and hue rotations (released as an augmented dataset of 886 materials),
        \changed{and its samples} render and relight as full BTFs.
\end{itemize}

\section{Related Work}
\label{sec:related}

\paragraph{Measured BTFs and neural materials.}
BTFs were introduced by Dana \etal~\cite{dana1999reflectance}; UBO2014~\cite{weinmann2014material} provides
84 measured materials in seven classes at 151$\times$151 light--view pairs and is distributed as a rank-limited
factorisation. Classic compression factorises the six-dimensional function linearly or
sparsely~\cite{kavoosighafi2023sparsebtf}. Neural approaches replaced the factorisation with learned decoders:
Rainer \etal~\cite{rainer2019neural} train a per-material autoencoder over per-texel angular profiles, and later a
unified encoder and decoder shared across materials~\cite{rainer2020unified}, which still stores a
32-dimensional latent per texel. NeuMIP~\cite{kuznetsov2021neumip} combines neural texture pyramids with learned
neural offsets for parallax, and follow-up work refines its architecture~\cite{xue2024hierarchical}, synthesises
BTFs of arbitrary extent from an example~\cite{xu2024dynamic,xu2025comprehensive}, and runs neural materials in
real-time path tracers~\cite{zeltner2024real}; Dou \etal~\cite{dou2024real} share neural primitives across
materials through a codebook. Neural Biplanes~\cite{fan2023neural} use a universal decoder with
per-material feature planes; NeuBTF~\cite{rodriguezpardo2023neubtf} conditions a neural BTF on a guidance image
for tiling and transfer; and Xia \etal~\cite{xia2026neural} fit multi-resolution neural textures with a neural
offset per material and extend a single material to larger extents by tiling its latent texture. Concurrently, RoboCloth~\cite{li2026robocloth} captures 500 real cloth materials and trains a shared decoder
with a dense latent texture per material. Idema and Peers~\cite{idema2024neural} infer per-pixel neural material parameters
from a single flash photograph. All of these store features per texel or per pixel. We compare against the
official implementations of~\cite{rainer2020unified,fan2023neural,xia2026neural}.

\paragraph{Latent codes, hypernetworks and generation in function space.}
Sztrajman \etal~\cite{sztrajman2021neural} represent a measured BRDF with a small MLP, an autoencoder gives a latent
space for editing measured BRDFs~\cite{hu2020deepbrdf}, and hypernetworks map a latent code to such networks and give
a space of BRDFs~\cite{gokbudak2024hypernetworks,li2024hypersnbrdf}, and physically based neural BRDFs enforce reciprocity and energy conservation for reconstruction and
editing~\cite{zhou2026pbnbrdf}. Fourier features~\cite{tancik2020fourier} allow coordinate networks
to fit high-frequency signals, and auto-decoders~\cite{park2019deepsdf} learn one code per shape for a shared
network. Generative models can then be trained on such neural fields: functa~\cite{dupont2022functa} fit a
generative model to per-datum modulations of a shared network, HyperDiffusion~\cite{erkoc2023hyperdiffusion}
diffuses directly in MLP weight space, and M$^3$ashy~\cite{zhou2026m3ashy} applies hyperdiffusion to neural fields
of measured, spatially uniform BRDFs. We extend the latent-code idea from spatially uniform BRDFs to spatially
varying, measured BTFs, where the decoder must also represent the spatial layout, and train the generative model
on one global code per material.

\paragraph{Generative material models.}
Most material generators produce SVBRDF maps for an analytic BRDF~\cite{kavoosighafi2024survey}.
MaterialGAN~\cite{guo2020materialgan} and TileGen~\cite{zhou2022tilegen} train StyleGAN generators on synthetic
SVBRDFs; MatFuse~\cite{vecchio2024matfuse}, ControlMat~\cite{vecchio2024controlmat}, Material
Palette~\cite{lopes2024material} and StableMaterials~\cite{vecchio2024stablematerials} use diffusion models
trained on synthetic material libraries such as MatSynth~\cite{vecchio2024matsynth}. Generated synthetic
materials often look synthetic, and several works learn from real photographs to close this gap: Henzler
\etal~\cite{henzler2021generative} learn a generative space of SVBRDF textures from flash photographs,
PhotoMat~\cite{zhou2023photomat} trains a generator of neural material features on 12,000 real flash photos, and
RealMat~\cite{zhou2025realmat} finetunes a diffusion model with a realism reward learned from real material
photographs. Beyond analytic BRDFs, a conditional GAN predicts BTF images of known materials under new light and view
directions~\cite{zhang2019btf}, MesoGAN~\cite{diolatzis2023mesogan} generates neural reflectance shells for
mesoscale appearance, Generative Neural Materials~\cite{raghavan2025generative} train a diffusion model over
neural material textures from a large corpus of synthetic materials, and video-diffusion approaches generate or
extract PBR and neural materials~\cite{munkberg2026texture,youwang2026extracting,xue2026videoneumat}. All of
these learn from synthetic materials, from single photographs, or from the priors of image and video models. We
learn from measured BTFs, which record the full light- and view-dependent appearance of real samples, including
parallax, masking and self-shadowing.

\paragraph{Scarce data, augmentation and memorisation.}
Measured BTFs are scarce: UBO2014 has 84 materials, and a concurrent capture has 500 cloth
samples~\cite{li2026robocloth}. Generative models reproduce their training data when it is small; even diffusion
models trained on large image sets copy training images~\cite{somepalli2023diffusion,carlini2023extracting}.
Augmentation is therefore common: M$^3$ashy~\cite{zhou2026m3ashy} enlarges MERL by permuting colour channels and
interpolating in PCA space, and Yu \etal~\cite{yu2026richer} procedurally enrich simple PBR materials with
layered, multi-lobe effects before training a generator. We instead recolour measured BTFs within their class
with a linear colour transfer~\cite{pitie2007linear}, so that every training value remains a measured value up to
a $3{\times}3$ colour transform, and the colourways keep the structure, relief and shadowing of the measured BTFs.

\section{Method}
\label{sec:method}

\begin{figure*}[t]
\centering
\colorlet{netgrey}{black!65}
\definecolor{encblue}{RGB}{207,224,243}\definecolor{nodelight}{RGB}{143,179,220}
\definecolor{nodemid}{RGB}{74,122,186}\definecolor{nodedark}{RGB}{38,72,118}
\begin{tikzpicture}[
    font=\footnotesize,
    arr/.style={-{Latex[length=1.6mm]}, semithick, black!70},
    img/.style={inner sep=0, draw=black!35, line width=0.3pt},
    code/.style={draw=black!60, fill=black!15, line width=0.4pt, minimum width=0.13cm, minimum height=0.9cm, inner sep=0},
    lab/.style={font=\footnotesize, align=center},
    note/.style={font=\scriptsize, black!75, align=center},
    panel/.style={draw=black!20, rounded corners=3pt, fill=black!2}]
  \newcommand{\NET}[6]{
    \begin{scope}[shift={(#1,#2)}]
      \pgfmathsetmacro\nl{dim(#3)}
      \pgfmathsetmacro\w{(\nl-1)*0.3+0.14}
      \pgfmathsetmacro\hz{#3[0]}
      \foreach \i in {1,...,\numexpr\nl-1\relax} {
        \pgfmathsetmacro\ha{#3[\i-1]} \pgfmathsetmacro\hb{#3[\i]}
        \fill[#4!12] ({(\i-1)*0.3+0.14},{-\ha/2}) -- ({\i*0.3},{-\hb/2}) -- ({\i*0.3},{\hb/2}) -- ({(\i-1)*0.3+0.14},{\ha/2}) -- cycle;
      }
      \foreach \i in {1,...,\nl} {
        \pgfmathsetmacro\h{#3[\i-1]}
        \filldraw[fill=#4!45, draw=#4!80!black, line width=0.35pt, rounded corners=0.7pt] ({(\i-1)*0.3},{-\h/2}) rectangle ++(0.14,\h);
      }
      \node[lab, above] at ({\w/2},{\hz/2+0.02}) {#5};
      \coordinate (#6-w) at (-0.1,0); \coordinate (#6-e) at ({\w+0.1},0); \coordinate (#6-s) at ({\w/2},{-\hz/2-0.05});
    \end{scope}}
  \newcommand{\TRAP}[9]{
    \begin{scope}[shift={(#1,#2)}]
      \def\dx{0.1}\def\dy{0.1}   %
      \filldraw[fill=encblue!55!white, draw=black!70, line width=0.5pt, line join=round] (0,{#4/2}) -- (#3,{#5/2}) -- ({#3+\dx},{#5/2+\dy}) -- (\dx,{#4/2+\dy}) -- cycle;
      \filldraw[fill=encblue!80!black, draw=black!70, line width=0.5pt, line join=round] (#3,{-#5/2}) -- ({#3+\dx},{-#5/2+\dy}) -- ({#3+\dx},{#5/2+\dy}) -- (#3,{#5/2}) -- cycle;
      \filldraw[fill=encblue, draw=black!70, line width=0.5pt, line join=round] (0,{-#4/2}) -- (#3,{-#5/2}) -- (#3,{#5/2}) -- (0,{#4/2}) -- cycle;
      \node[font=#8, align=center, inner sep=0] at ({#3/2},#9) {#6};
      \coordinate (#7-w) at (-0.04,0); \coordinate (#7-e) at ({#3+\dx+0.04},0); \coordinate (#7-s) at ({#3/2},{-(#4+#5)/4-0.04});
    \end{scope}}
  \newcommand{\MLPC}[2]{
    \begin{scope}[shift={(#1)}]
      \filldraw[fill=white, draw=black!70, line width=0.5pt, rounded corners=5pt] (-0.17,-0.46) rectangle (1.05,0.46);
      \foreach \l/\n in {0/1,1/4,2/4,3/4,4/1}
        \foreach \k in {1,...,\n}
          \coordinate (q-\l-\k) at (\l*0.22, {(\k-(\n+1)/2)*0.2});
      \foreach \l/\n/\m in {0/1/4,1/4/4,2/4/4,3/4/1} {
        \pgfmathtruncatemacro\lp{\l+1}
        \foreach \a in {1,...,\n} \foreach \b in {1,...,\m}
          \draw[black!50, line width=0.35pt] (q-\l-\a) -- (q-\lp-\b);
      }
      \foreach \l/\n/\c in {0/1/nodelight,1/4/nodemid,2/4/nodemid,3/4/nodemid,4/1/nodedark}
        \foreach \k in {1,...,\n}
          \filldraw[fill=\c, draw=white, line width=0.4pt] (q-\l-\k) circle (0.062);
    \end{scope}}

  \node[img] at ({0.75+0.16*2},{0.16*2}) {\includegraphics[width=1.15cm]{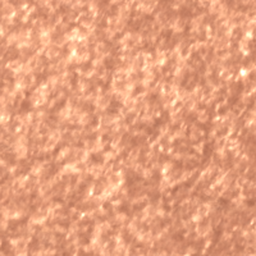}};
  \node[img] at ({0.75+0.16*1},{0.16*1}) {\includegraphics[width=1.15cm]{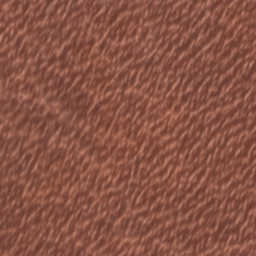}};
  \node[img] at ({0.75+0.16*0},{0.16*0}) {\includegraphics[width=1.15cm]{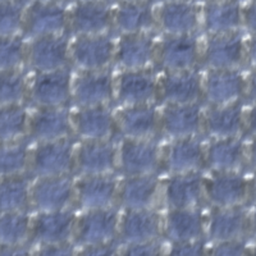}};
  \node[note, below] at (0.91,-0.72) {84 measured BTFs};
  \node[font=\small] at (2.0,0.08) {$+$};
  \node[img] at ({2.95+0.16*2},{0.16*2}) {\includegraphics[width=1.15cm]{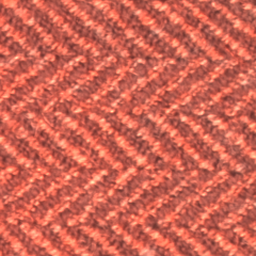}};
  \node[img] at ({2.95+0.16*1},{0.16*1}) {\includegraphics[width=1.15cm]{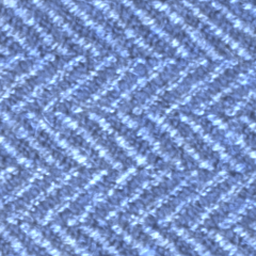}};
  \node[img] at ({2.95+0.16*0},{0.16*0}) {\includegraphics[width=1.15cm]{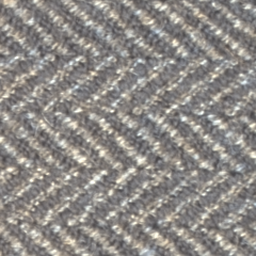}};
  \node[note, below] at (3.11,-0.72) {colourways};

  \TRAP{4.1}{0.1}{1.25}{1.7}{0.8}{Set\\Encoder}{enc}{\footnotesize}{0.06}
  \draw[arr] (3.85,0.1) -- (enc-w);
  \foreach \k in {2,1,0}
    \node[code] at ({5.9+0.1*\k},{0.03+0.1*\k}) {};
  \node[note, below] at (6.0,-0.47) {codes $\mathbf{z}_m$};
  \draw[arr] (enc-e) -- (5.75,0.1);

  \node[note] at (2.22,-2.35) {noise $\boldsymbol\epsilon\sim\mathcal{N}(0,I)$\\class $c$};
  \begin{scope}[shift={(3.95,-2.35)}]
    \begin{scope}[shift={(0.525,0)}, scale=1.25, shift={(-0.525,0)}]   %
    \filldraw[fill=encblue, draw=black!70, line width=0.5pt, rounded corners=4pt] (0,-0.55) rectangle (1.05,0.55);
    \draw[-{Latex[length=1.2mm]}, black!90, thick] (0.12,-0.35) .. controls (0.45,-0.45) and (0.55,0.35) .. (0.93,0.33);
    \draw[-{Latex[length=1.2mm]}, black!60, thick] (0.12,0.0) .. controls (0.4,-0.15) and (0.6,0.35) .. (0.93,0.2);
    \draw[-{Latex[length=1.2mm]}, black!35, thick] (0.12,0.35) .. controls (0.4,0.3) and (0.6,0.1) .. (0.93,0.07);
    \end{scope}
    \node[lab, above] at (0.525,0.66) {rectified flow $v_\phi$};
  \end{scope}
  \draw[arr] (3.38,-2.35) -- (3.75,-2.35);
  \node[code] (zh) at (6.0,-2.35) {};
  \node[note, below] at (6.0,-2.8) {sampled code $\hat{\mathbf{z}}$};
  \draw[arr] (5.22,-2.35) -- (zh.west);
  \draw[black!35, dashed, -{Latex[length=1.4mm]}] (6.0,-0.95) -- (6.0,-1.85);

  \TRAP{7.3}{-1.12}{1.6}{1.0}{2.0}{Shared\\Decoder\\$g_\theta$}{dec}{\footnotesize}{-0.1}
  \draw[arr] (6.27,0.1) -- (6.95,0.1) |- ($(dec-w)+(0,0.35)$);
  \draw[arr] (6.27,-2.35) -- (6.95,-2.35) |- ($(dec-w)+(0,-0.35)$);
  \draw[arr] ($(dec-s)+(0,-0.45)$) node[below, inner sep=1pt] {$\mathbf{x},\boldsymbol\omega_i,\boldsymbol\omega_o$} -- (dec-s);

  \draw[arr] (dec-e) -- (9.25,-1.12) |- (9.75,0.1);
  \draw[arr] (dec-e) -- (9.25,-1.12) |- (9.75,-2.35);
  \node[img] at (10.4,0.1) {\includegraphics[width=1.2cm]{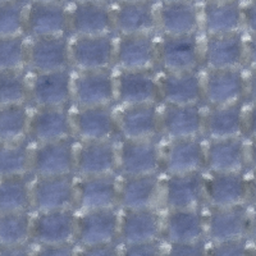}};
  \node[img] at (11.7,0.1) {\includegraphics[width=1.2cm]{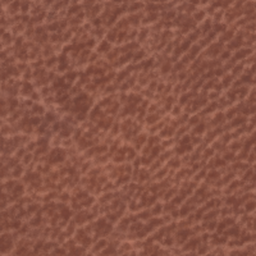}};
  \node[note, below] at (11.05,-0.55) {reconstructed};
  \node[img] at (10.4,-2.35) {\includegraphics[width=1.2cm]{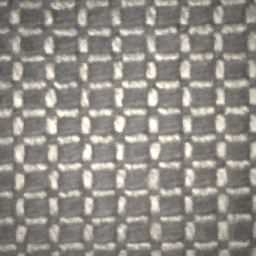}};
  \node[img] at (11.7,-2.35) {\includegraphics[width=1.2cm]{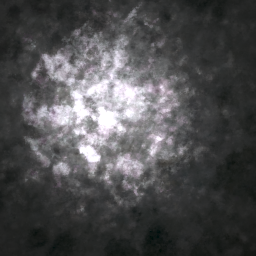}};
  \node[img] at (13.0,-2.35) {\includegraphics[width=1.2cm]{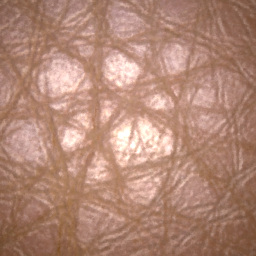}};
  \node[note, below] at (11.7,-3.0) {samples};
  \draw[black!40, dashed] (13.65,-2.35) -- (13.95,-2.35);
  \node[img] at (14.6,-2.35) {\includegraphics[width=1.2cm]{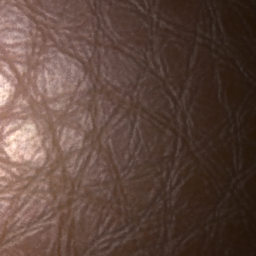}};
  \node[img] at (15.9,-2.35) {\includegraphics[width=1.2cm]{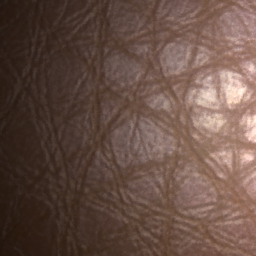}};
  \node[note, below] at (15.25,-3.0) {relit: light left, right};

  \node[inner sep=0, anchor=west] at (12.500,0.1) {\includegraphics[height=1.30cm]{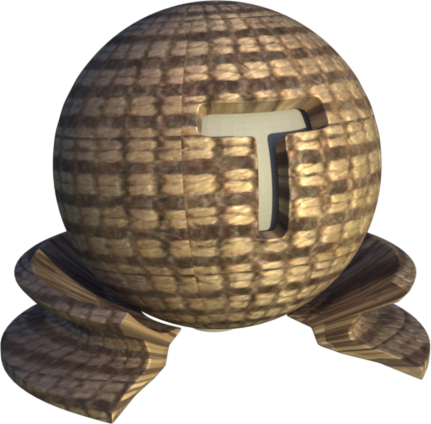}};
  \node[inner sep=0, anchor=west] at (14.296,0.1) {\includegraphics[height=1.30cm]{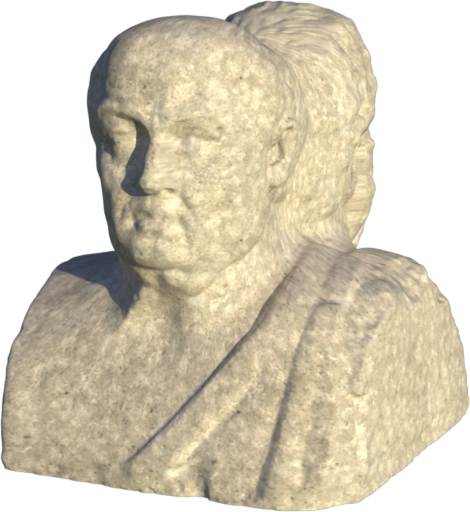}};
  \node[inner sep=0, anchor=west] at (15.964,0.1) {\includegraphics[height=1.30cm]{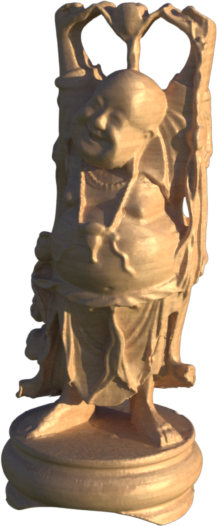}};
  \node[note, below] at (14.50,-0.6) {rendered on objects};
\end{tikzpicture}
\caption{\textbf{Overview.} The compact model encodes each measured material and its colourways (\cref{sec:aug})
into one 256-float code with a set encoder; a decoder $g_\theta$ shared by all materials maps the code, texel position
and light and view directions to RGB. A class-conditional rectified flow fitted to these codes samples codes,
which the same decoder renders as full BTFs that can be relit. Top right: \changed{measured materials reconstructed by the compact model, rendered on objects.}
}
\label{fig:overview}
\end{figure*}

A BTF $B(\mathbf{x}, \boldsymbol{\omega}_i, \boldsymbol{\omega}_o)$ gives the reflected RGB radiance factor at
texel $\mathbf{x}$ of a planar sample for light direction $\boldsymbol{\omega}_i$ and view direction
$\boldsymbol{\omega}_o$. Unlike an SVBRDF, it is measured rather than modelled, and therefore includes the
sample's meso-scale effects as they appear at each texel. We parameterise the texel position by $\mathbf{x} \in [-1,1]^2$, and the direction pair by the Rusinkiewicz half and difference vectors
$(\mathbf{h}, \mathbf{d}) \in \mathbb{R}^6$~\cite{rusinkiewicz1998new}, keeping the azimuth because BTF texels are anisotropic.
\Cref{fig:overview} summarises our approach.

\subsection{Per-material BTF}
\label{sec:permat}
\changed{We first build a per-material representation: a Fourier-feature MLP fitted to a single material, which gives our most
accurate results and later becomes the backbone of our compact model (\cref{sec:repr}). For each material $m$, we fit}
\begin{equation}
  \hat{B}_m(\mathbf{x}, \boldsymbol{\omega}_i, \boldsymbol{\omega}_o) =
  \exp\!\big(g_m([\gamma(\mathbf{x}), \mathbf{h}, \mathbf{d}])\big) - 1,
  \label{eq:permat}
\end{equation}
where $\gamma(\mathbf{x}) = [\sin(2\pi \mathbf{F}\mathbf{x}), \cos(2\pi \mathbf{F}\mathbf{x})]$ are random Fourier
features~\cite{tancik2020fourier} with $\mathbf{F} \in \mathbb{R}^{128 \times 2}$ drawn from $\mathcal{N}(0, 16^2)$,
and $g_m$ is an MLP with four hidden layers of width 1024 and ReLU activations (3.42M parameters). The
exponential output follows neural BRDF representations~\cite{sztrajman2021neural} and keeps predictions in the
dynamic range of measured reflectance.

We minimise a log-space $L_1$ loss on cosine-weighted values over $n$ random queries
$(\mathbf{x}, \boldsymbol{\omega}_i, \boldsymbol{\omega}_o)$ drawn from all texels and all measured direction pairs,
\begin{equation}
  \mathcal{L} = \frac{1}{n}\sum \big|\log(1 + \cos\theta_i\,\hat{B}) - \log(1 + \cos\theta_i\,B)\big|.
  \label{eq:loss}
\end{equation}
The logarithm balances specular highlights against dark, shadowed texels.

\subsection{Compact model}
\label{sec:repr}
The compact model replaces the per-material network by one shared decoder $g_\theta$ and a single code
$\mathbf{z}_m \in \mathbb{R}^{256}$ per material,
\begin{equation}
  \hat{B}_m(\mathbf{x}, \boldsymbol{\omega}_i, \boldsymbol{\omega}_o) =
  \exp\!\big(g_\theta([\gamma(\mathbf{x}), \mathbf{h}, \mathbf{d}, \mathbf{z}_m])\big) - 1,
  \label{eq:decoder}
\end{equation}
with 512 Fourier frequencies drawn from $\mathcal{N}(0, 32^2)$ and four hidden layers of width 3072 (32.3M parameters,
shared by all materials). We condition the decoder by concatenating the material code with each query.

\paragraph{Encoding.}
A permutation-invariant set encoder~\cite{zaheer2017deep} infers $\mathbf{z}_m$ from a set of $k$ measured
samples $\{(\mathbf{x}_j, \mathbf{h}_j, \mathbf{d}_j, \mathbf{b}_j)\}_{j=1}^{k}$ with measured RGB values
$\mathbf{b}_j$, applying an MLP $f_\psi$ to each sample and averaging,
\begin{equation}
  \mathbf{z}_m = \frac{1}{k}\sum_{j=1}^{k} f_\psi(\mathbf{x}_j, \mathbf{h}_j, \mathbf{d}_j, \mathbf{b}_j).
\end{equation}
The encoder maps a set of measured samples, each consisting of a spatial position, a light--view pair, and its
corresponding colour, to a compact latent code. The decoder then uses this code, together with a spatial position
and a light--view pair, to predict the corresponding colour. During training we draw
a fresh random context of $k{=}16{,}384$ samples per material and step; at test time we
encode from all samples. Rendering only requires $\mathbf{z}_m$ and $g_\theta$.

\paragraph{Training.}
Each step samples four materials and, for each, $n{=}65{,}536$ random queries, and minimises \cref{eq:loss} plus
$\lambda\,\lVert \mathbf{z}_m \rVert_2^2$ with $\lambda{=}10^{-2}$; the penalty keeps the codes small and on a common scale, so that
the code space stays compact for the generative model. A single UBO2014 material contains
$400^2 \times 151^2 \times 3 \approx 1.1 \times 10^{10}$ values; we train for 705k steps with cosine learning-rate
decay, so each measured sample is seen only 0.6 times on average.

\subsection{Colourways}
\label{sec:aug}
\begin{figure}[t]
  \centering
  \includegraphics[width=\linewidth]{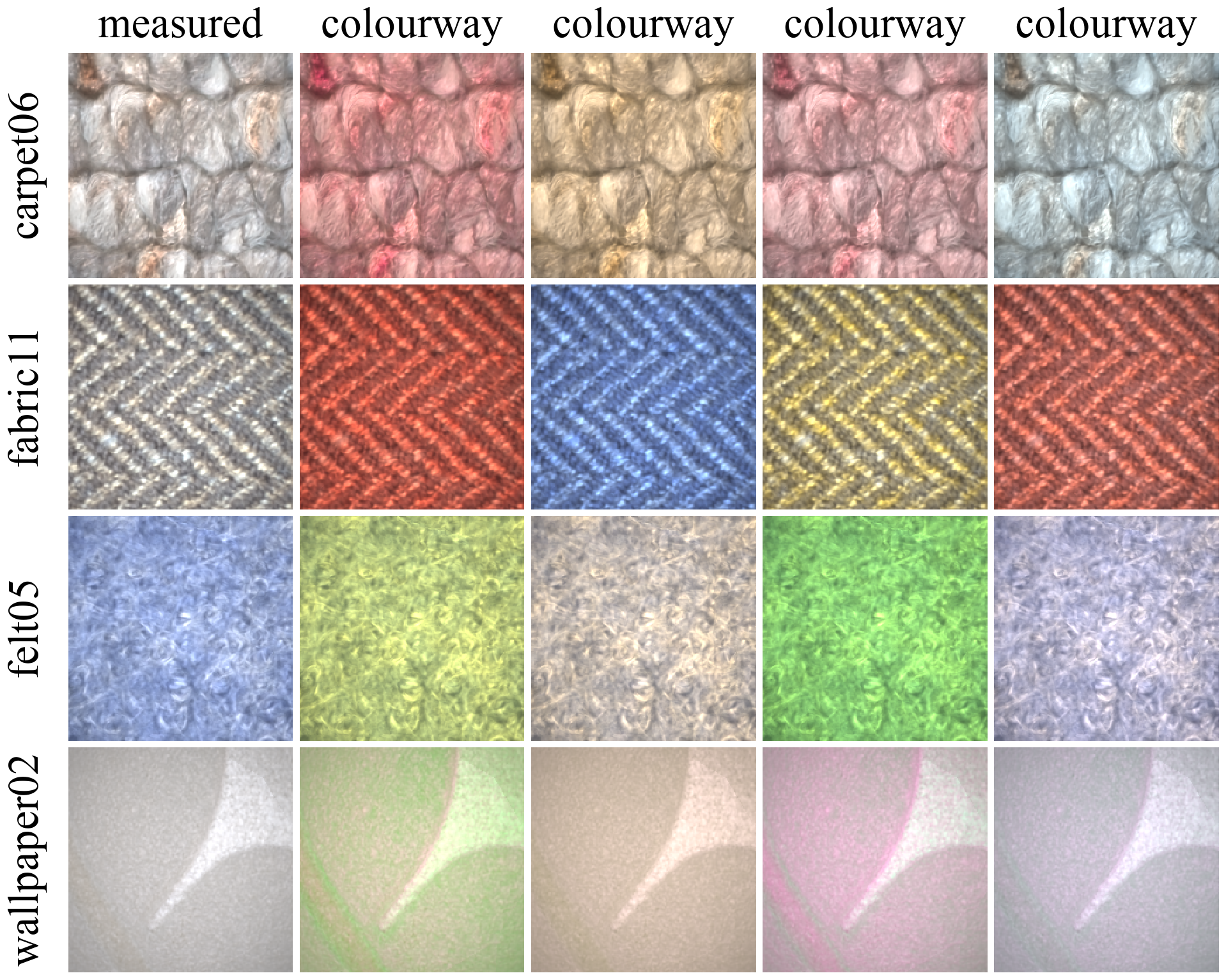}
  \caption{\textbf{Colourways.} One measured material per row (left) and four of its colourways: its structure in the
  colours of other materials of its class, as flash photographs. Every value is a measured value under one $3{\times}3$
  colour transform.}
  \label{fig:colourways}
\end{figure}

UBO2014 has only twelve materials per class, which makes it difficult to learn a distribution of materials. We
therefore enlarge the 84 measured materials with \emph{colourways} (\cref{fig:colourways}): for two materials $A$ and $B$ of the same
class, we recolour $A$ with the colours of $B$. \changed{Since a BTF stores RGB reflectance rather than colour, one $3{\times}3$
matrix $M_{AB}$ is applied to every measured RGB reflectance value of $A$,} at every texel and every light--view pair,
so the result keeps the structure, relief and shadowing of $A$ while its colours change. \changed{Negative values, which the
matrix can produce, are set to zero; values above one are kept, as the reflectance of a measured BTF can go above one.} This mirrors how fabrics, leathers and wallpapers are produced in colourways: the same
weave, grain or print in different dyes. We fit $M_{AB}$ by linear colour transfer~\cite{pitie2007linear}, \changed{mapping the mean and covariance of the colours in
$A$'s frontal image onto those of $B$'s. In total, the 84 measured materials yield 1{,}141 training materials, including
hue rotations for felt and carpet (details in supplementary Sec.~S3).}

For generation, the compact model is finetuned on the augmented training set, starting from an earlier checkpoint
(supplementary Sec.~S1).

\subsection{Generation}
\label{sec:gen}
We learn the distribution of codes with a class-conditional rectified flow~\cite{liu2022flow,lipman2023flow}.
Given normalised codes $\mathbf{z}_1$ of training materials, \changed{their class labels $c\in\{1,\dots,7\}$, the
index of their UBO2014 class (carpet, fabric, felt, leather, stone, wallpaper or wood),} and Gaussian noise
$\mathbf{z}_0$, the velocity network $v_\phi$ is trained with
\begin{equation}
  \mathcal{L}_{\text{flow}} = \mathbb{E}_{t,\mathbf{z}_0,\mathbf{z}_1}
  \big\lVert v_\phi\big((1-t)\mathbf{z}_0 + t\mathbf{z}_1,\, t,\, c\big) - (\mathbf{z}_1 - \mathbf{z}_0) \big\rVert^2,
\end{equation}
dropping the class with probability 0.1 for classifier-free guidance~\cite{ho2022classifier}. $v_\phi$ is an MLP of width 512
with four residual blocks (4.1M parameters) whose scale, shift and gate are modulated by \changed{the time embedding and a learned embedding of the class label} (adaptive
layer normalisation). The flow is trained on the codes of \changed{886 materials, leaving out the 255 colourways and hue rotations whose
texture is faint} (supplementary Sec.~S3). At test time we integrate
$\dot{\mathbf{z}} = v_u + w\,(v_c - v_u)$ from noise, where $v_c$ and $v_u$ are the velocities with and without the
class, with 100 Euler steps and guidance weight $w{=}2$, and decode
the resulting code with $g_\theta$ into a full BTF that can be evaluated at any texel, light and view.

\section{Experiments}
\label{sec:exp}

\paragraph{Data.}
We use all 84 materials of UBO2014~\cite{weinmann2014material}, seven classes of materials found in indoor scenes
(carpet, fabric, felt, leather, stone, wallpaper, wood) with twelve material instances each. Every material is a
5\,cm$\times$5\,cm planar sample captured fully automatically under directional illumination as 22{,}801 HDR images,
one for every combination of 151 view and 151 light directions. The images are resampled to $400\times400$ texels and
distributed as a rank-101 factorisation per material; we decode this factorisation and use it as the reference
\changed{for training and evaluating all methods}.

\paragraph{Protocol.}
All methods are evaluated on nine materials covering every class (carpet01, fabric01/05/07, felt01, leather01,
stone01, wallpaper01, wood01) with a single shared protocol, \changed{whose results are reported in \cref{tab:recon}: each method predicts the BTF
directly at eight light--view pairs spread evenly over the measured grid, from frontal to the grazing limit of the
capture (light and view at $75^\circ$ elevation), and is compared with the measured images at the same pairs};
one exposure per material, shared by all methods; PSNR, SSIM and
LPIPS~\cite{zhang2018unreasonable} on tonemapped images; and $\Delta E_{\mathrm{ITP}}$ on linear HDR values (shown to correlate well with subjective quality assessment \cite{Kavoosighaﬁ_2025_BRDF_quality}).
Baselines are run with the authors' official code and, where available, official pretrained weights; we only
convert our data to their input format. As in the evaluation of Rainer \etal~\cite{rainer2020unified}, the
evaluated materials are part of every method's training set.

\paragraph{Implementation.}
All models are trained in PyTorch with Adam on a single NVIDIA A100; the compact model uses bfloat16 mixed precision,
with the Fourier features in float32. Each per-material network is trained for 30k steps with batches of 32{,}768 queries and a learning rate of $5\times10^{-4}$.
The compact model is trained for 705k steps with a learning rate of at most $5\times10^{-4}$ and cosine decay. The flow is trained for 8k steps. Further training details are in supplementary Sec.~S1.

\subsection{Reconstruction}
\label{sec:exp_recon}

\begin{table}[t]
\centering
\small
\resizebox{\linewidth}{!}{%
\begin{tabular}{lrcccc}
\toprule
Method & Floats/mat. & PSNR$\uparrow$ & SSIM$\uparrow$ & LPIPS$\downarrow$ & $\Delta E\downarrow$ \\
\midrule
\multicolumn{6}{l}{\emph{Per-material fitting}} \\
GGX, per-texel anisotropic & 1.76M & 22.10 & 0.703 & 0.234 & 24.80 \\
NeuMIP~\cite{kuznetsov2021neumip} & 2.62M & 28.34 & 0.782 & 0.192 & 11.19 \\
Xia \etal~\cite{xia2026neural} & 3.32M & 33.24 & 0.898 & 0.084 & 6.53 \\
Ours, per-material & 3.42M & \textbf{36.44} & \textbf{0.928} & \textbf{0.060} & \textbf{4.54} \\
\midrule
\multicolumn{6}{l}{\emph{Shared model across materials}} \\
Neural Biplanes~\cite{fan2023neural} & 2.88M & 26.16 & 0.639 & 0.355 & 15.65 \\
Rainer \etal, $d{=}8$~\cite{rainer2020unified} & 1.60M & 29.90 & 0.792 & 0.171 & 9.77 \\
Rainer \etal, $d{=}32$~\cite{rainer2020unified} & 5.44M & 31.37 & \textbf{0.862} & \textbf{0.132} & 8.67 \\
Ours, compact & \textbf{256} & \textbf{32.56} & 0.820 & 0.183 & \textbf{6.77} \\
\bottomrule
\end{tabular}}
\caption{\textbf{Reconstruction on nine UBO2014 materials.} Floats/mat.: numbers stored per material, excluding
decoders shared by all materials.
Even counting shared decoders, amortised over the 84 materials, the compact model stores 4--14$\times$ fewer values
than Rainer \etal ($\approx$385k vs.\ 1.6M and 5.44M). Rainer \etal store per texel a $d$-dimensional latent plus the mean and standard deviation used to normalise its values.}
\label{tab:recon}
\end{table}

\begin{figure}[t]
  \centering
  \includegraphics[width=\linewidth]{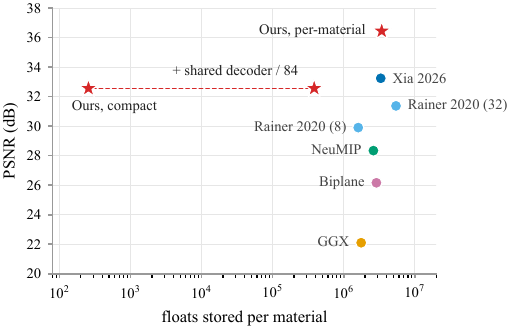}
  \caption{\textbf{Rate--distortion} on nine UBO2014 materials: PSNR against the number of values stored per
  material (log scale). The compact model is shown with its 256-float code and with the shared decoder amortised
  over 84 materials.}
  \label{fig:rd}
\end{figure}

\begin{figure*}[t]
  \centering
  \includegraphics[width=\linewidth]{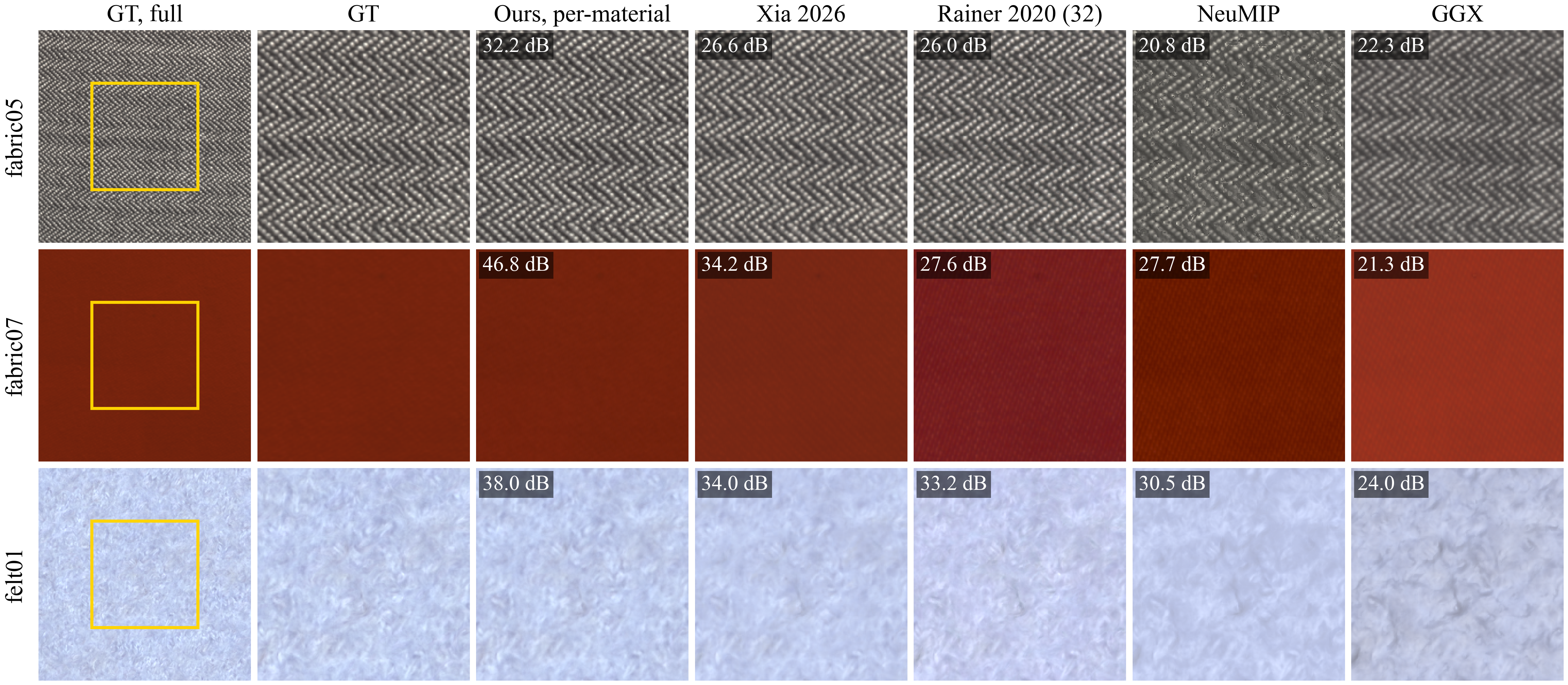}
  \caption{\textbf{Per-material reconstruction.} First column: the full measured image, with the crop shown for
  every method outlined; same light--view pair and exposure for all methods in a row; PSNR of the full image. The compact model is compared qualitatively in supplementary Sec.~S2.
}
  \label{fig:recon}
\end{figure*}

\Cref{tab:recon} groups methods by whether a network is fitted to each material or shared across materials.

\paragraph{Per-material fitting.}
Our per-material network outperforms the recent neural BTF of Xia \etal~\cite{xia2026neural} by 3.2\,dB
PSNR at matched size (3.42M vs.\ 3.32M parameters), with better SSIM, LPIPS and colour error on every one of the
nine materials (\cref{fig:recon}). A per-texel
anisotropic GGX fit~\cite{walter2007microfacet}, which can express any material produced by metallic--roughness PBR
generators, is 14\,dB below our network. As a per-texel BRDF, it \changed{treats each} texel in isolation and cannot
represent the parallax, masking and shadowing between neighbouring texels that a BTF captures~\cite{dana1999reflectance}.

\paragraph{Shared models.}
Our compact model reaches a higher PSNR and lower colour error than Rainer \etal~\cite{rainer2020unified} with
32-dimensional per-texel latents, which shift the colours of the measured materials
(supplementary Sec.~S2). Even counting its 32.3M-parameter shared decoder, amortised over the 84
materials, it stores about 385k values per material, 14$\times$ fewer than their 5.44M; the per-material part is a
256-float code. Per-texel latents keep a lower LPIPS (0.132 vs.\ 0.183), but at 14$\times$ the storage per material (\cref{fig:rd}).
The compact model is the only method at the kilobyte scale: a whole measured BTF, about $10^{10}$ values, is described
by a single 256-float code, with one decoder shared by all materials. Every material thus becomes a single point in a
common space from which materials can be sampled (\cref{sec:exp_gen}), whereas a per-texel representation remains an
image of millions of features per material. Qualitative comparisons of the compact model are in supplementary Sec.~S2.

\subsection{Generation}
\label{sec:exp_gen}

\begin{figure*}[t]
  \centering
  \includegraphics[width=\linewidth]{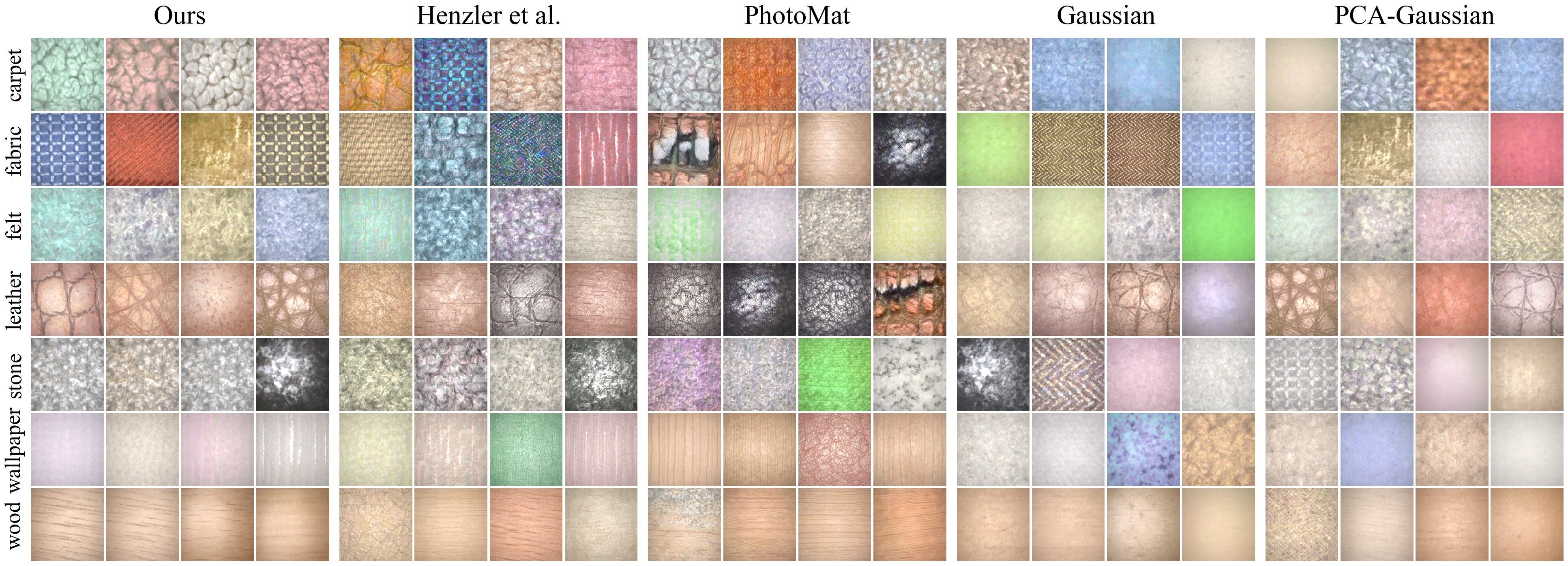}
  \caption{\textbf{Generated materials as flash photographs}: four random samples per class of each generator,
  without selection; shown is the central half of each photograph, the region used for KID. The Gaussian baselines
  sample our code space and are decoded by our decoder. Henzler \etal are retrained on our 886 images and sample each class from a Gaussian over the latent codes of its training images.
  PhotoMat is finetuned from its released model and is unconditional, so we group its samples by the class of their
  nearest training image in DINOv2 features.}
  \label{fig:gen_compare}
\end{figure*}

\paragraph{Baselines and setup.}
\changed{No previous generator learns from measured BTFs; the closest material generators learn from flash
photographs~\cite{henzler2021generative,zhou2023photomat}, which can be captured in large numbers, and are therefore the
natural image-based baselines for our flow. As they cannot be compared with ours in BTF space, we compare all
generators as flash photographs:} the training materials are rendered
from their BTFs with a point light at the camera, in the capture geometry of Henzler \etal (field of view
$45^\circ$, $384\times384$ pixels, gamma 2.2); these 886 images are the training data of the image-based baselines and
the reference for all metrics. Our samples and those of the Gaussian baselines are rendered by our decoder in the same
geometry.
\changed{To test whether our code space needs a flow-based density estimator, we compare with established density models over
the same codes, which also share our decoder:} a per-class Gaussian with full covariance fitted to the flow's training
codes, and the same Gaussian restricted to the top 16 principal directions. We retrain Henzler \etal~\cite{henzler2021generative}
with their official code and default settings (15 epochs) on the 886 training images. PhotoMat~\cite{zhou2023photomat} is trained on thousands of real
flash photographs; we finetune their released $256\times256$ generator on our 886 images with their training code. We render its samples with the light at the camera.

\paragraph{Qualitative comparison.}
\Cref{fig:gen_compare} shows random samples of every generator, without selection.
\changed{Ours and the Gaussian baselines are
sampled for each class directly. Neither image-based generator is trained with classes: Henzler \etal encode each
training image, so we sample each class from a Gaussian over the codes of its images, while PhotoMat has no encoder,
so each of its samples is assigned the class of its nearest training image in DINOv2 features.} Our samples keep the structures
of the measured materials of their class: the loops of carpet, the grids and twills of fabric, the cells of leather,
the grain of wood.
\changed{Henzler \etal give sharp textures with more variety within a class and the highest coverage, as their encoder builds
on an ImageNet-pretrained network, but as analytic SVBRDFs they cannot represent the self-shadowing and parallax of
measured BTFs.
PhotoMat, finetuned on our 886 images, produces realistic wood, leather, felt and stone in places,
but collapses to a few modes and breaks down elsewhere: wood-like planks appear in the wallpaper and fabric rows, and
several fabric and leather samples are broken, high-contrast patches.} The Gaussian baselines sample everywhere around the class mean, but most of their samples land between
materials: decoded, they are blurry or show the structure of another class, such as fabric weaves in stone.

\begin{figure}[t]
  \centering
  \includegraphics[width=\linewidth,trim=0 9.5 0 6,clip]{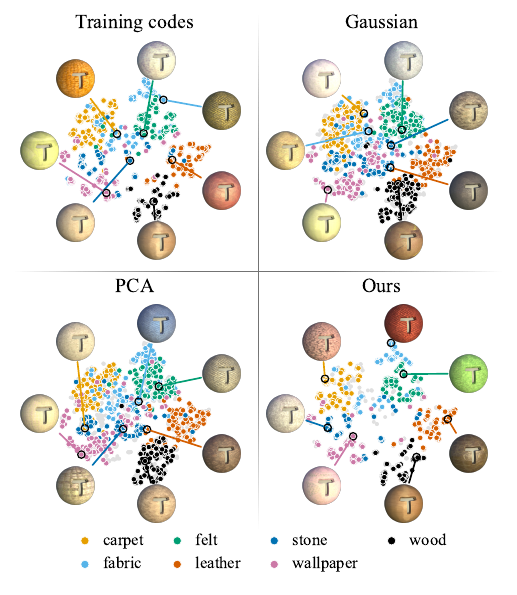}
  \caption{\textbf{Generated codes.} t-SNE of the training codes (measured materials and colourways) and of the 672
  samples of each class-conditional generator in our code space, coloured by the class they were generated for. The first
  panel shows the training codes by class; in the others they are shown in grey. Thumbnails show one random sample per class\changed{, rendered on a test object}. \changed{Henzler \etal and PhotoMat are
  not included, as they do not generate codes in our space.}}
  \label{fig:tsne}
\end{figure}

\begin{figure}[t]
  \centering
  \includegraphics[width=\linewidth]{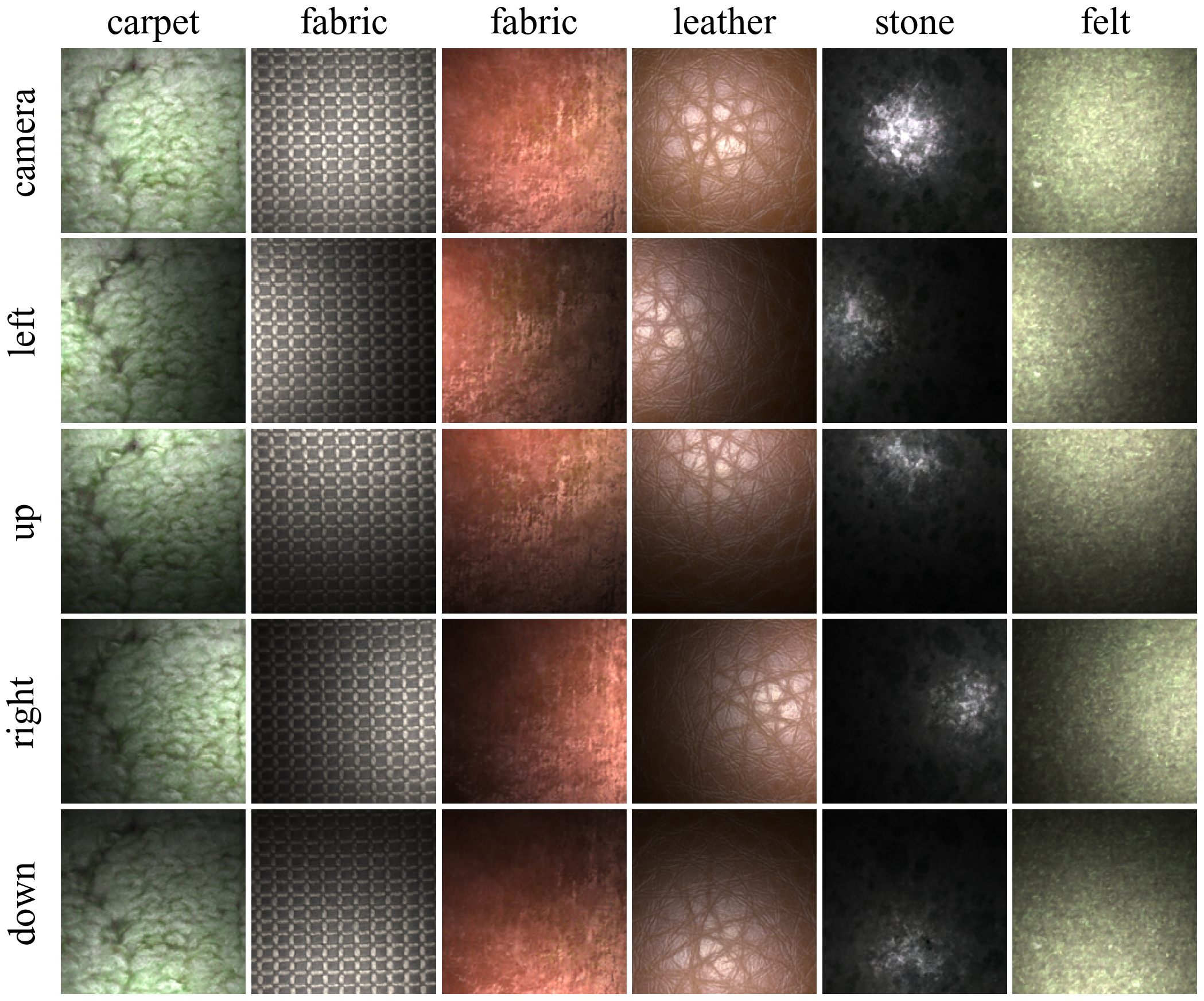}
  \caption{\textbf{Relighting generated materials.} Each column is one generated sample, rendered with a fixed
  camera and a point light at the camera (top row) and moved left, up, right and down (rows 2--5). Exposure is fixed
  per column.}
  \label{fig:relight}
\end{figure}

\Cref{fig:tsne} shows this in our code space: Gaussian and PCA samples fall between the classes and mix their
characteristics, while \changed{ours remain within their own class and share its characteristics}.
Because every sample is a full BTF, it can be relit (\cref{fig:relight} and supplementary Sec.~S6: highlights, shadows and
relief follow the light as in the measured data). Rendering a sample needs a one-off decoding into a light--view table
(3.7\,min on one A100; supplementary Sec.~S7); after that, it costs the same as a tabulated measured BTF.

\begin{table}[t]
\centering
\small
\setlength{\tabcolsep}{5pt}
\begin{tabular}{lccc}
\toprule
Generator & KID$\downarrow$ & MMD$\downarrow$ & COV$\uparrow$ \\
\midrule
Gaussian & 1222\,$\pm$\,61 & 0.276 & 18.6 \\
PCA-Gaussian & 1160\,$\pm$\,40 & 0.265 & 21.1 \\
Henzler \etal~\cite{henzler2021generative} & \phantom{0}698\,$\pm$\,20 & 0.252 & \textbf{23.0} \\
PhotoMat~\cite{zhou2023photomat} & \phantom{0}863\,$\pm$\,29 & 0.341 & 20.2 \\
\midrule
Ours, 84 measured & \underline{\phantom{0}656\,$\pm$\,44} & \underline{0.242} & 10.8 \\
Ours, with colourways & \textbf{\phantom{0}466\,$\pm$\,37} & \textbf{0.215} & \underline{22.1} \\
\bottomrule
\end{tabular}
\caption{\textbf{Generation on UBO2014}: 672 samples per generator as flash photographs against the flash
photographs of the 886 training materials, in DINOv2 features. KID$\times10^3$, $\pm$ s.d.\ over ten
random halves; COV in \%. Best in bold, second best underlined. All generators are trained on the same 886
materials, except the first row of ours; PhotoMat is finetuned from its released model.}
\label{tab:gen}
\end{table}

\paragraph{Quantitative comparison.}
We draw 96 samples per class (672 in total) and compare them with the 886 training images in DINOv2
features~\cite{oquab2024dinov2,stein2023exposing}: KID~\cite{binkowski2018demystifying} measures the distance between
the two distributions, MMD how closely the samples match the training images, and coverage (COV) their
diversity~\cite{achlioptas2018learning}. All images are compared on their central half, at the scale of PhotoMat's
samples (supplementary Sec.~S4). Our flow has the lowest KID and MMD (\cref{tab:gen}), and decoded training materials,
the best our codes allow, reach a KID of 372.

\subsection{Ablations}
\label{sec:ablation}

\begin{figure}[t]
  \centering
  \includegraphics[width=\linewidth]{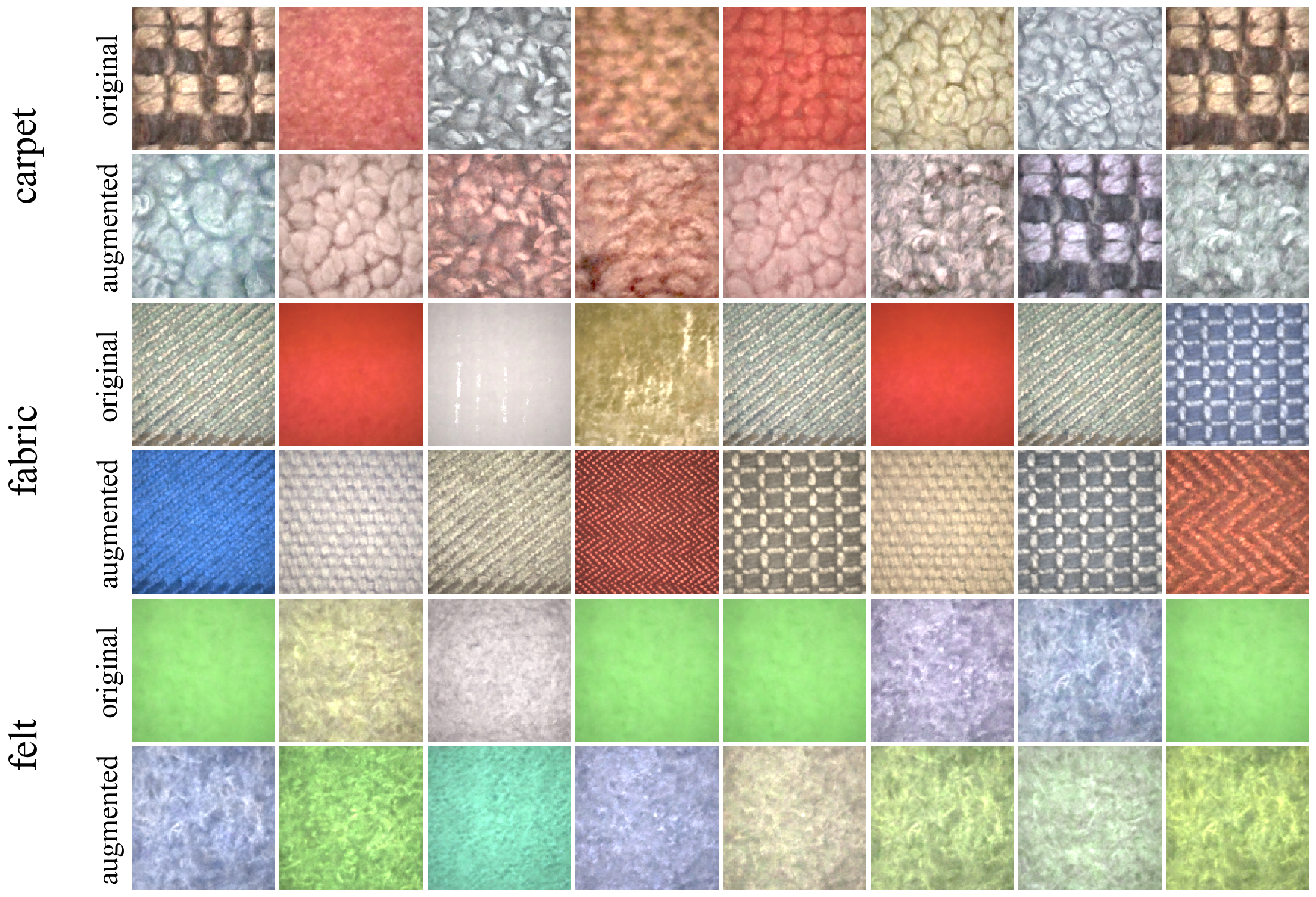}
  \caption{\textbf{Colourways ablation}: eight random samples per class of the flow trained on the 84 measured
  materials only (original) and with colourways (augmented), as flash photographs. Without colourways the flow
  repeats a few training materials.}
  \label{fig:ablation_colourways}
\end{figure}

\paragraph{Colourways.}
Trained on the 84 measured materials only, the same flow copies a few training materials
(\cref{fig:ablation_colourways}). With colourways, KID improves from 656 to 466 and coverage doubles from 10.8\% to
22.1\% (\cref{tab:gen}): the generator covers many more colours of the measured structures.

\subsection{Discussion}
\label{sec:discussion}

\paragraph{Global codes versus per-texel latents.}
Per-texel latent maps~\cite{rainer2020unified,xia2026neural} and feature planes~\cite{fan2023neural} store most of a
material's information explicitly and use a small decoder; our compact representation stores almost none and uses a
large shared decoder. For reconstruction this trades perceptual sharpness for far smaller
per-material storage. For generation, a per-texel representation is an image of 8--32 features per texel, 1.6--5.4M values
per material, and generating one requires an image-scale generative model. A global code reduces each material to a
point in a 256-dimensional space, where a small flow models their distribution.

\paragraph{Extent and tiling.}
Our decoder covers the measured patch. Xia \etal~\cite{xia2026neural} extend a single material to larger extents with
a diffusion model over its latent textures, which is complementary to \changed{our generation and could extend it}.

\paragraph{Limitations.}
With twelve measured materials per class, our generator learns the structures it has seen: its samples stay close
to the measured materials and their colourways, and structural novelty is limited to a few blends of two training
materials (supplementary Sec.~S5). Also, rendering requires decoding each material once into \changed{a light--view table}.

\section{Conclusion}
\label{sec:conclusion}

We presented neural representations of measured BTFs at two operating points: a per-material network that
represents a measured BTF accurately, and a compact model that reduces each material to a single
code. In this code space, measured BTFs become sampleable: a class-conditional flow, trained on the measured
materials enlarged with colourways and hue rotations, generates materials that render and relight as full BTFs. This is a first step towards bringing the
realism of measured appearance to generated materials in rendering pipelines. Larger captures such as
RoboCloth~\cite{li2026robocloth}, augmentation that varies the structure of materials as well as their colour, and conditioning on text prompts or images instead of classes are natural next
steps. We hope this
work makes measured appearance a practical target for generative modelling.

{
    \small
    \bibliographystyle{ieeenat_fullname}
    \bibliography{draft/refs}
}
\end{document}